\documentclass[aps,epsfig]{revtex4} %eqsecnum,
\usepackage[dvips]{graphicx} 
\usepackage{amssymb}

\newcommand{\rf}[1]{\ref{fig:#1}}

\newcommand{\be}{\begin{equation}}
\newcommand{\ee}{\end{equation}}
\newcommand{\bea}{\begin{eqnarray}}
\newcommand{\eea}{\end{eqnarray}}
\newcommand{\eql}[1]{\label{eq:#1}}       
\newcommand{\ec}[1]{equation~(\ref{eq:#1})}          
\newcommand{\refeq}[1]{equation~(\ref{eq:#1})}          
\newcommand{\Ec}[1]{~(\ref{eq:#1})}

\newcommand{\phidot}{\dot{\Phi}}

\newcommand{\phibar}{\bar{\phi}}
\newcommand{\phibardot}{\dot{\bar{\phi}}}

\newcommand{\alphadot}{\dot{\alpha}}

\newcommand{\adotova}{\frac{\dot{a}}{a}}
\newcommand{\bdotovb}{\frac{\dot{b}}{b}}
\newcommand{\ex}{e^{-4 \bar{\phi}}} 

\newcommand{\beq}{\begin{equation}}
\newcommand{\eeq}{\end{equation}}
\newcommand{\beqarr}{\begin{eqnarray}}
\newcommand{\eeqarr}{\end{eqnarray}}

\renewcommand{\v}[1]{\mathbf{#1}}
\newcommand{\Deg}{^{\circ}}

\begin{document}

\title{Galaxy-CMB Cross-Correlation as a Probe of Alternative Models of
Gravity}

\author{Fabian Schmidt$^{1,2}$, Michele Liguori$^3$, Scott Dodelson$^{4,1,2}$}

\affiliation{$^1$Department of Astronomy \& Astrophysics, The University of Chicago, 
Chicago, IL~~60637-1433}
\affiliation{$^2$Kavli Institute for Cosmological Physics, 
Chicago, IL~~60637-1433}
\affiliation{$^3$Department of Applied Mathematics and Theoretical Physics, Centre for
Mathematical Sciences, University of Cambridge, Wilberfoce Road, Cambridge, CB3 0WA, United
Kingdom}
\affiliation{$^4$Center for Particle Astrophysics,
Fermi National Accelerator Laboratory, Batavia, IL~~60510-0500}

\date{\today}

%\twocolumn

\begin{abstract}
Bekenstein's alternative to general relativity, TeVeS, reduces to Modified
Newtonian Dynamics (MOND) in the galactic limit. On cosmological scales,
the (potential well$\leftrightarrow$overdensity) relationship is quite different than in
standard general relativity. Here we investigate the possibility of cross-correlating
galaxies with the cosmic microwave background (CMB) to probe this relationship. 
At redshifts of order
2, the sign of the CMB-galaxy correlation differs in TeVeS from that in general
relativity. We show that this effect is detectable and hence can serve as a
powerful discriminator of these two models of gravity.
\end{abstract}

\maketitle

\section{Introduction}

The standard theory of gravity, general relativity (GR),
coupled with the standard model of particle physics cannot account for many
cosmological observations, ranging from the dynamics of stars in galaxies to the
large scale expansion of the universe. This discrepency may well be
resolved by adding to the standard model of particle physics. For example, dark
matter in the form of supersymmetric particles could account for flat rotation
curves and the large scale structure in the universe. Dark energy in the form of
a cosmological constant or a scalar field could explain the acceleration of the
universe. This scenario is commonly called the ``$\Lambda$CDM model''.

Alternatively, general relativity may be wrong. Perhaps the theory needs to be
modified on large scales to account for cosmological observations. Many attempts
have been made along these
lines~\cite{Carroll:2004de,Easson:2004fq,Cognola:2005de,
Woodard:2006nt,Carloni:2004kp,Amarzguioui:2005zq,Nojiri:2006ri}, to explain either
(or sometimes both) the acceleration of the universe
and the phenomena usually associated with dark matter. One such attempt is
MOdified Newtonian Dynamics (MOND) initially proposed~\cite{Milgrom:1983pn} 
to explain flat rotation
curves in galaxies. Bekenstein~\cite{Bekenstein} has recently introduced TeVeS, a theory which
reduces to MOND in the appropriate limit but is robust enough to make
predictions in many different arenas.

Several groups~\cite{Skordis:2005xk,Skordis:2005eu,Dodelson:2006zt,Bourliot:2006ig} 
have now studied the evolution of cosmic structure in TeVeS and the
theory seems to be holding up. In some senses, this is
surprising~\cite{Lue:2003if} since any no-dark
matter theory faces the significant hurdle of explaining why the density field
has gone nonlinear. Observations of the cosmic microwave background (CMB) 
pin down the baryon inhomogeneities
at recombination to one part in $10^{4}-10^5$, and in standard general
relativity, these inhomogeneities grow by only a factor of a thousand from
that epoch until today. 

As shown in \cite{Dodelson:2006zt}, overdensities in TeVeS grow faster than in
standard general relativity (GR). The mechanics of this is subtle: the vector field
in TeVeS develops an instability and this growth in turn sources a difference
between the two Newtonian scalar potentials. The two potentials
$\Phi$ and $\Psi$ characterize scalar perturbations to the metric:
\be
ds^2 = -dt^2\left[ 1+ 2\Psi\right] + a^2(t)d{\bf x}^2 \left[ 1 + 2\Phi\right]
.\eql{metric}\ee
In $\Lambda$CDM with standard GR, these two potentials
are nearly equal to one another, so the TeVeS situation of unequal potentials is
unfamiliar territory. In particular, the relation between gravitational
potential and overdensity is quite different than in standard GR.
Fig.~\rf{poissonratio}
illustrates this. 

\begin{figure} [t]
\begin{center}
\includegraphics[width = .5\textwidth]{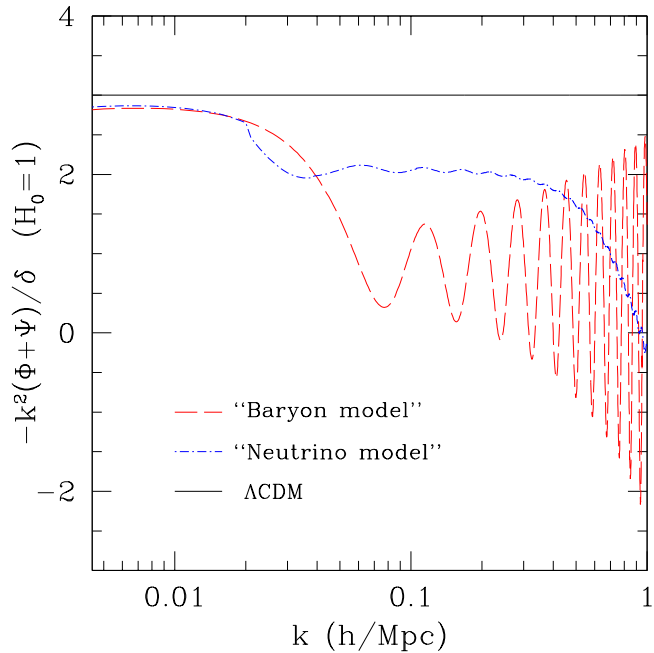}
\caption{The ``poisson ratio''  
$ {-k^2 (\Phi+\Psi)/(\Omega_b \delta_b + \Omega_\nu \delta_\nu)} $ 
for the two TeVeS models considered in section {\ref{sec:poisson}}. 
In these units, the Poisson
ratio is constant and equal to 3 in standard $\Lambda$CDM cosmology 
(solid line). 
This is true in TeVeS only for the largest
scales, where the standard term in the Poisson equation dominates. On
smaller scales the vector field perturbations 
dominate and the Poisson ratio exhibits a much more complicated 
behavior (see \S \ref{sec:poisson} for more details).}
\label{fig:poissonratio}
\end{center}
\end{figure}

In this paper, we demonstrate that one can
test this prediction of TeVeS (for a different test, see ~\cite{Zhang:2007nk})
by cross-correlating a deep galaxy survey with
the CMB. A number of groups have already succeeded in
cross-correlating tracers of the density field with the 
CMB~\cite{Scranton:2003in,Afshordi:2003xu,Fosalba:2003ge,Padmanabhan:2004fy}. 
The non-zero correlation is a result of the time-dependent
gravitational potentials at late times because the universe is not completely
matter dominated. In TeVeS, the potentials are determined not only by the matter
overdensities but also by the behavior of the vector field. So the pattern of
the cross-correlation changes. Most strikingly, at redshift of order 2$-$3 the
cross-correlation changes sign. Observations of such a sign change would be a
smoking gun signature of a modification of gravity.

Before proceeding with the calculation, we note that we focus on TeVeS as a
fairly well-defined test model. Bertschinger~\cite{Bertschinger:2006aw} 
has pointed out however that
generic modifications of gravity lead to changes of the sort discussed here.
So we expect the cross-correlations envisioned here to be powerful tools
cosmologists will use to discriminate (general relativity + dark matter + dark
energy) from many alternative theories of gravity. This idea has
already been applied in the context of DGP and $f(R)$ gravity for which
specific
CMB-galaxy cross-correlation signatures at high redshifts 
have been predicted by the authors of~\cite{Song:2006jk,Zhang05}. 
Recently, more general parametrizations of modifications to gravity
have been proposed \cite{Amendola,Caldwell}. These authors concentrated
on modifications of gravity intended to explain late-time acceleration. However,
TeVeS, with the aim of explaining the breadth of cosmological observations 
without dark matter, shows much more severe deviations from general relativity.
In particular, the post-Newtonian parameter introduced in \cite{Caldwell}
is applicable if it is close to unity, which is not the case for TeVeS 
throughout the whole history of the universe. Hence, it is preferable in this
case to compare the theory directly with observations.

\section{Integrated Sachs-Wolfe Effect}\label{sec:ISW}

A CMB photon traveling through time-varying gravitational potentials 
$\Phi$ and $\Psi$ in the metric\Ec{metric} will suffer a change in energy,
resulting in a change in the observed CMB temperature: the so-called
Integrated Sachs-Wolfe (ISW) effect (\cite{SachsWolfe,Kofman,CrittendenTurok}, see \cite{Afshordi} for a review). The fractional change in CMB temperature
in the direction $\v{\hat{n}}$ is given by an integral along the line of sight back
to the redshift of last scattering :
\be
\frac{\Delta T(\v{\hat{n}})}{T} = \int_0^{z_{\rm LSS}} dz\; \frac{d}{dz}\left ( \Phi(\v{r},\:z) + \Psi(\v{r},\:z) 
\right ) |_{\v{r} = \chi(z)\v{\hat{n}}}
\ee
where $\chi(z)$ is the comoving distance out to redshift $z$. The evolution of the gravitational 
potentials therefore determines the ISW effect.
On the other hand, 
the fractional overdensity of galaxies along the line of sight is
\be
\delta_g(\v{\hat n}) = \int_0^\infty dz W(z) \delta_g(\chi(z) \v{\hat n};z)
\ee
where $W(z)$ is the redshift distribution of galaxies (normalized so it integrates
to unity), determined both 
by the intrinsic
redshift distribution of galaxies and by the properties of the survey at hand.

Since we expect an overdensity of galaxies to be correlated with potential wells, we
expect a non-zero correlation between the observed temperature anisotropy and the
observed angular galaxy distribution. 
This correlation is however model-dependent and can be used to distinguish
between different theories of gravity.

If we focus on large
scales, the fractional galaxy overdensity in Fourier space is
\be
\delta_g(\v{k}, z) = b(z) \delta(\v{k}, z=0) D(k,z)
\ee
where $b(z)$ is the time-dependent but scale-independent bias, and $D(k,z)$
is the linear growth function, normalized to unity today. In standard $\Lambda$CDM, the growth function
depends on $z$ only, but more generally (and in TeVeS in particular) it can have
a scale dependence. With this restriction to large angular scales, the
angular cross-power spectrum at a multipole $l$ is
\be
C_{gT}(l) = \frac{2}{\pi} \int_0^\infty dk\:k^2\:P(k) \:\delta_g(l,k)\:\delta_{\rm ISW}(l,k)
\label{eq:CgTexact}
\ee
where $P(k)$ is the matter power spectrum today, and the two weighting
functions for galaxies and the ISW effect are:
\bea
\delta_g(l,k) &=& \int_0^\infty dz\: W(z)\:b(z) \:D(z,k)\:j_l(k\,\chi(z)) \\
\delta_{\rm ISW}(l,k) &=& \int_0^{z_{\bf LSS}} dz \frac{d}{dz} \left( - {\Phi(k, z) + \Psi(k,z) \over
 \delta(k, z=0)} \right)\cdot j_l(k\,\chi(z)) \equiv \int_0^{z_{\bf LSS}} dz\: D_{\rm ISW}(k,z) j_l(k\,\chi(z))
\eql{disw}\eea
Here, $\delta_g(l,k)$ gives the contribution of the modes with wave number $k$ to 
the projected overdensity of galaxies for the
spherical harmonic $l$ and $j_l$ is the spherical Bessel function. 

On small angular scales, i.e. large $l \gtrsim 10$, the Limber approximation can be used:
\be
\int dk\: k^2\: f(k) \: j_l(k\,\chi) \:j_l(k\,\chi') \rightarrow \frac{\pi}{2}\frac{1}{\chi^2} \:f(k=(l+1/2)/\chi) \:\delta_{\rm Dirac}(\chi-\chi'),
\ee
where $f(k)$ is a slowly varying function. This simplifies the expression 
for the galaxy-CMB cross-power coefficients to a single integral 
(we set $c = 1$):
\be
C_{gT}(l) = \int dz \frac{H(z)}{\chi^2(z)} 
D(z,k_{\perp})\:b(z)\:W(z)\: D_{\rm ISW}(k_{\perp}, z)\: 
P(k_{\perp}), \quad k_{\perp} \equiv \frac{l+1/2}{\chi(z)}.
\eql{iswint}
\ee
Once $C_{gT}(l)$ is calculated, the galaxy-CMB angular correlation function 
$w_{gT}(\theta)$ is given by (e.g., \cite{LoVerde:2006cj}):
\be
\label{eq:wgT}
w_{gT}(\theta) = \sum_{l=2}^{\infty} \frac{2l + 1}{4\pi} C_{gT}(l)
\ee
We use the exact expression (\refeq{CgTexact} and following) up to $l=10$. 
For larger $l$,
we use the Limber approximation, as the differences to the exact calculation
are very small, about $10^{-3}$, as we explicitly verified.

\begin{figure}[t]
\begin{minipage}[ht]{8.3cm}
\begin{center}
\includegraphics[width=8.3cm,clip]{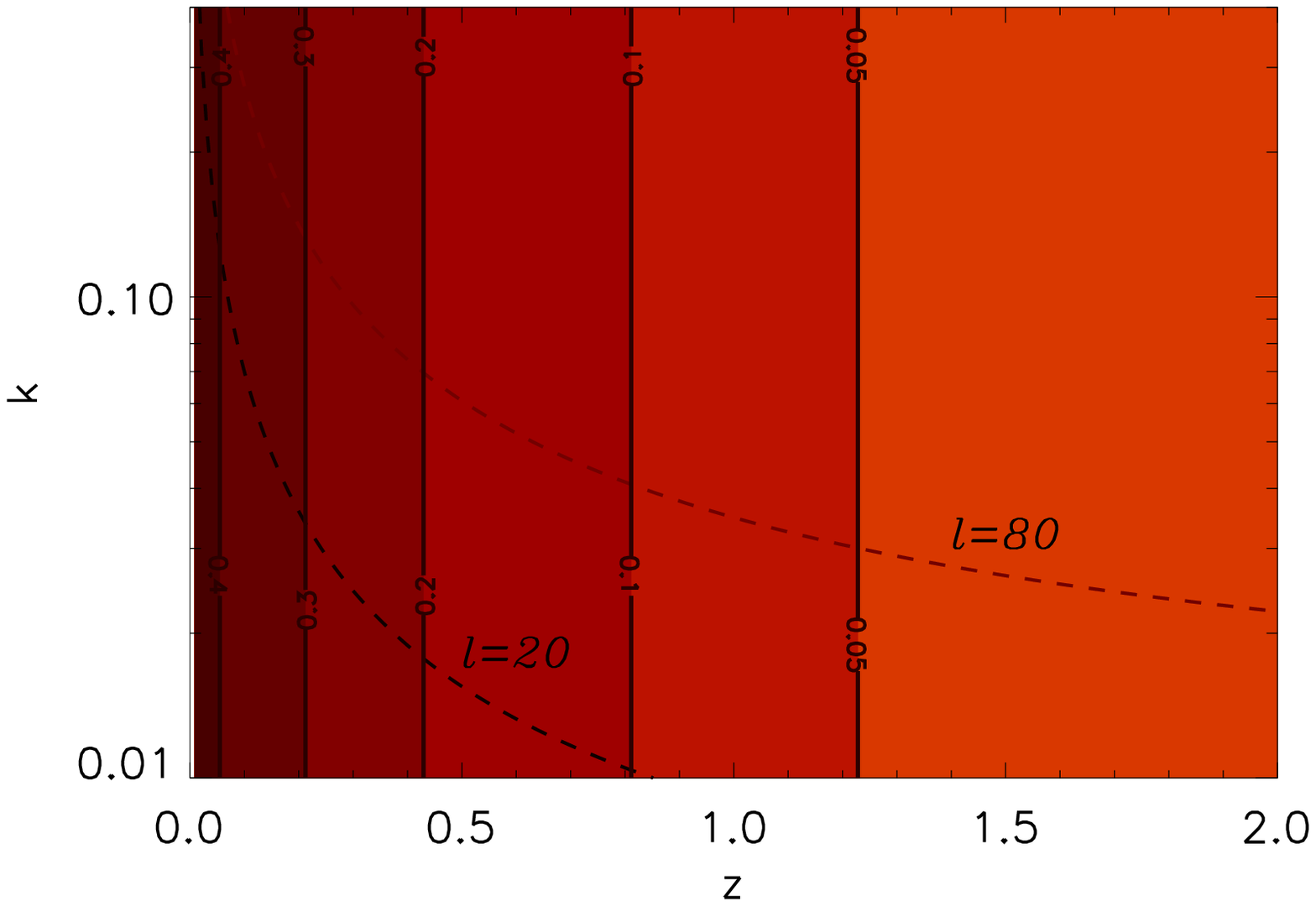}
\end{center}
\end{minipage}
\hfill
\begin{minipage}[ht]{8.3cm}
\begin{center}
\includegraphics[width=8.3cm,clip]{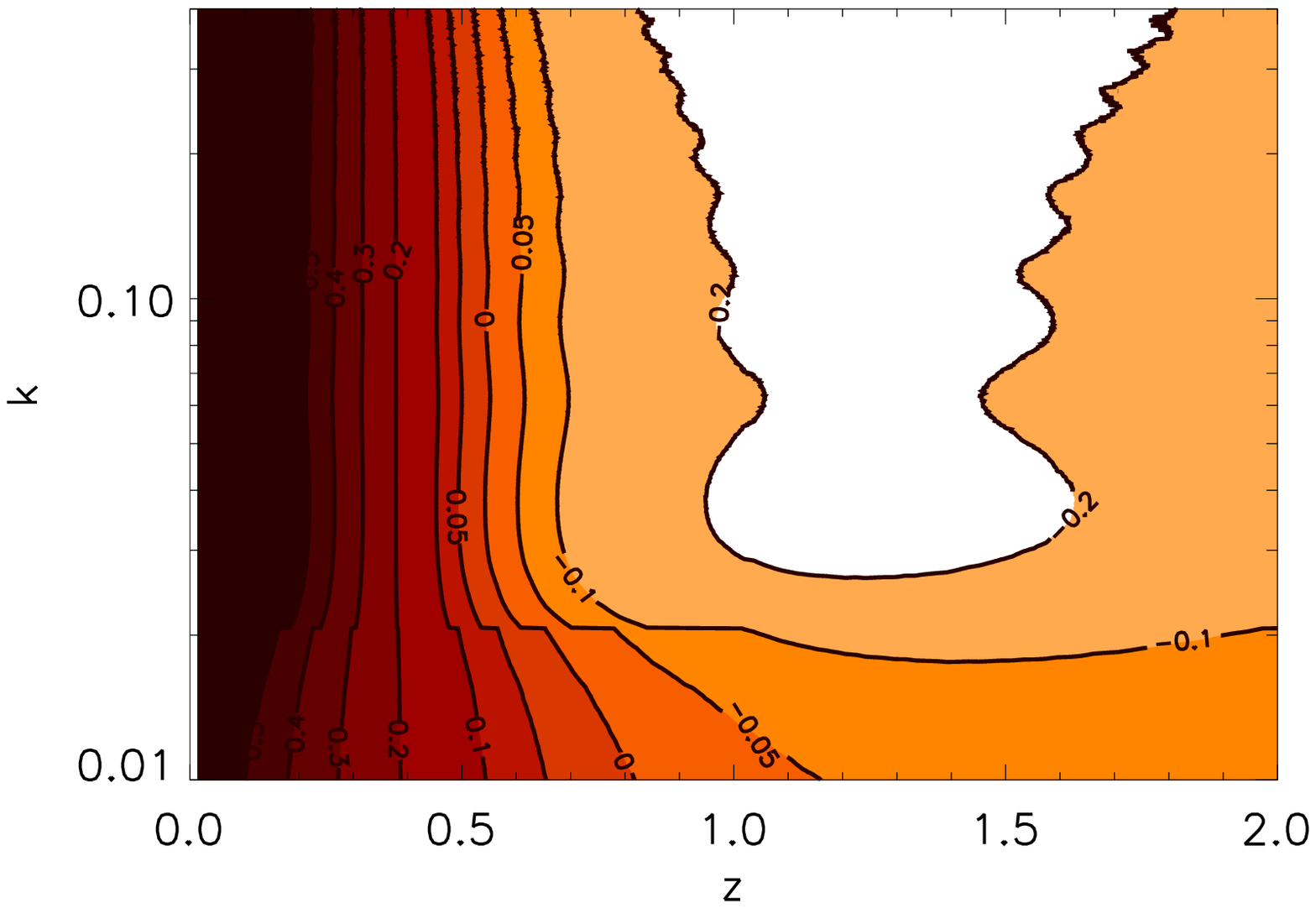}
\end{center}
\end{minipage}
\caption{\label{fig:iswlcdm} {\it Left Panel.} 
$k^2D_{\rm ISW}/H_0^2$ as a function of wavenumber $k$ and redshift in
standard $\Lambda$CDM. Here we show results from linear theory so the relevant
growth functions do not depend on wavenumber. Dashed curves show the line along which the
integral is performed for two values of $l$.
{\it Right panel.}  $k^2D_{\rm ISW}/H_0^2$ as a function of wavenumber $k$ and redshift in
TeVeS. Note that at high reshift, the TeVeS
weighting function goes negative.
}
\end{figure}

Equation\Ec{CgTexact} and following, or equivalently \refeq{iswint}, state 
that the $C_{gT}(l)$ are determined by 
a weighted integral of $D_{\rm ISW}(k,z)$ along the line 
$k=(l+1/2)/\chi(z)$ in the $(k,z)$-plane. 
The function $D_{\rm ISW}$ defined in \ec{disw} represents the response of the 
potentials to the growth of matter overdensities. In the case of
standard $\Lambda$CDM, it is given by the Poisson equation:
\be\label{eq:GRpoisson}
k^2 \Phi = \frac{3}{2} H_0^2\: \Omega_m (1+z) \delta(k,z),\;\mbox{and}\;
\Psi = \Phi
\;\Rightarrow\; D_{\rm ISW}(k,z) = -\frac{3}{k^2} H_0^2\: \Omega_m\; \frac{d}{dz} (1+z) D(z)
\ee
The left panel of Fig.~\rf{iswlcdm} 
shows $D_{\rm ISW}$ as a function of $k$ and $z$ in the standard $\Lambda$CDM
model and the right panel for
TeVeS. The dashed curves in the left panel show the line along which
the integral is performed in $(k,z)$ space for two values of $l$. Larger values of $l$
probe smaller scales and therefore higher $k$-values.
At low $z$, the two theories make similar predictions 
(the larger amplitude for TeVeS at low redshift is model-dependent), whereas the 
differences become large for redshifts larger than one. 

The striking differences in between the two panels in Fig.~\rf{iswlcdm} are 
due to the differences in the growth of structure in TeVeS and standard 
$\Lambda$CDM during the matter dominated era. In absence of cold
dark matter, the growth factor in TeVeS must be enhanced with respect to 
$\Lambda$CDM in order to match the present amplitude of perturbations. 
This enhancement produces an evolution of the
gravitational potential, and thus an ISW signal during matter
domination. This does not happen in $\Lambda$CDM, where the
ISW contribution comes {\em only} from the decay of the potential
during dark energy domination.

At redshifts of order one, $D_{\rm ISW}$ even becomes negative 
for TeVeS, leading to negative cross-power.
For large z samples, this feature of TeVeS should have observable 
consequences. Before examining these consequences, we
explain the features of TeVeS that lead to the behavior depicted in
Fig.~\rf{iswlcdm}, in the following section.

\section{The Generalized Poisson equation in TeVeS}\label{sec:poisson}

Perturbations around the smooth cosmology in TeVeS were first studied in 
\cite{Skordis:2005xk} with the full set of evolution equations derived in
\cite{Skordis:2005eu}. The equations which determine the scalar potentials
of \ec{metric} are~\cite{Skordis:2005eu}:
\bea
\label{eq:potential1}
-2 k^2 \Phi - 2 e^{4 \phibar} \bdotovb \left(3\dot{\Phi} + k^2 \zeta +
3 \bdotovb \Psi \right) &=&
K_B k^2 E + 8 \pi G
a^2 \bar\rho \delta \\
\label{eq:potential2}
\phidot + \bdotovb \Psi &=& 
4 \pi G a^2 e^{-4 \phibar} (\rho + P) \Theta 
\eea
where $\phibar$ is the zero order smooth part of the TeVeS scalar field; $b\equiv ae^{\phibar}$;
$\alpha$ and $E$ characterize perturbations to the vector field; $\bar\rho$ is
the background matter density; $\Theta$ is proportional to the
velocity of the matter;
\be
\zeta\equiv(e^{-4\phibar}-1)\alpha;
\eql{zetadef}\ee
and $K_B$ is a dimensionless parameter governing the 
importance of the vector field. Here we have neglected the scalar 
field perturbations as these are small.

In the following we will assume matter domination, including baryons
and massive neutrinos (no cold dark matter). 
Using equation~(\ref{eq:potential2}) we
can write the term in parenthesis in equation~(\ref{eq:potential1}) as:
\beq
\left(3\dot{\Phi} + k^2 \zeta + 3 \bdotovb \Psi \right) =
12 \pi G a^2 e^{-4 \phibar} (\rho + P) \Theta + k^2 \zeta  .
\eeq
On sub-horizon scales we can neglect the $\Theta$ term. Then,
using the definition of $\zeta$
and $\bdotovb = \adotova + \phibardot$, we obtain the generalized Poisson equation: 
\beq\label{eq:poisson}
\Phi = -\frac{3H_0^2\left(\Omega_b \delta_b + \Omega_\nu \delta_\nu
  \right)}{2a k^2} - \frac{K_B E}{2} -\left(\adotova + \phibardot \right) 
 \left(1 - e^{4 \phibar} \right)\alpha \; .
\eeq
The first term on the right is the standard contribution from matter, but the last two are
unique to TeVeS. Perturbations to the vector field affect the behavior of the gravitational potential,
destroying the simple 1-1 relationship between potential and density implicit in general relativity.

\begin{figure}[t]
\begin{minipage}[ht]{0.45\textwidth}
\begin{center}
\includegraphics[width=\textwidth,clip]{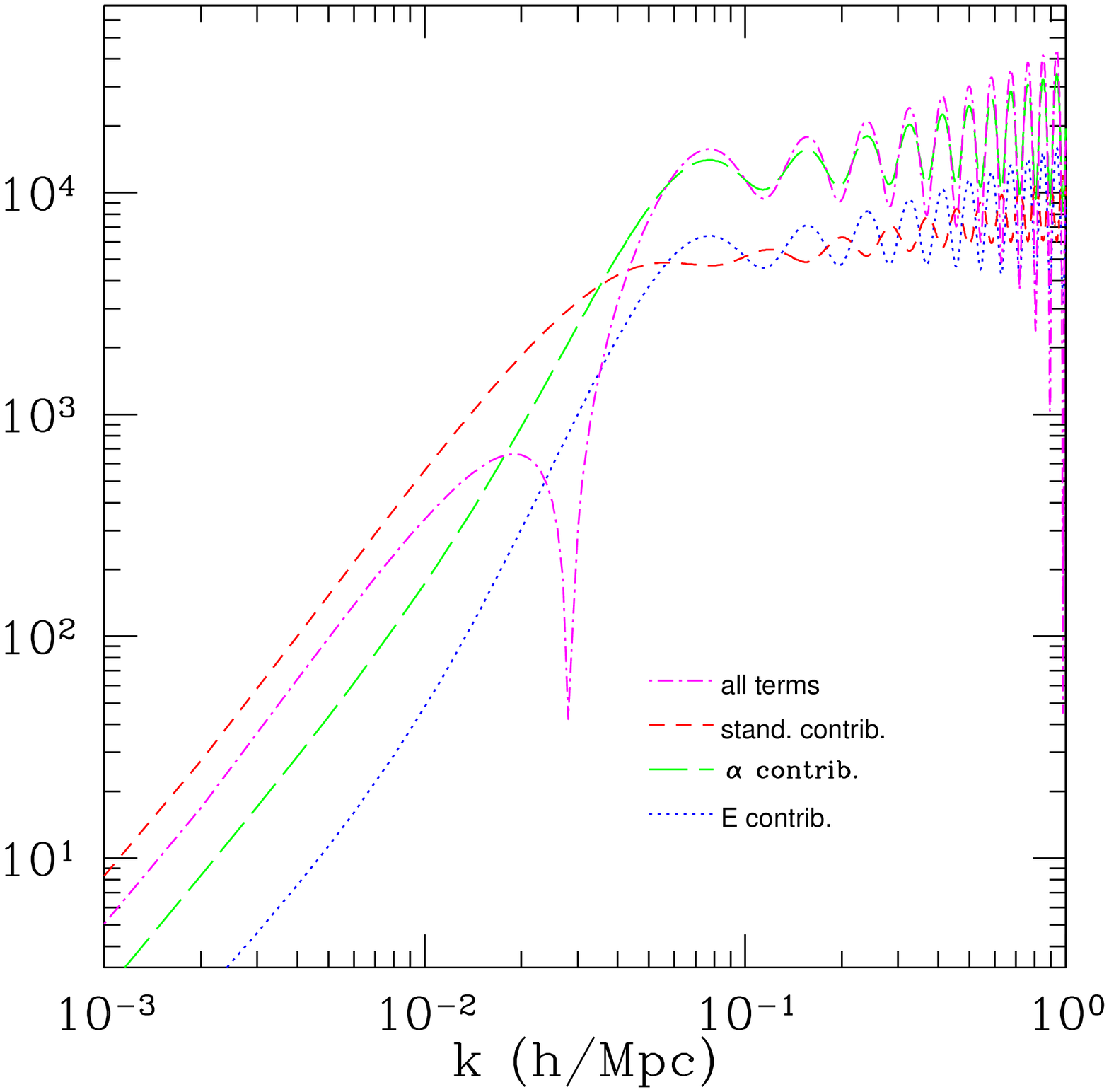}
\end{center}
\end{minipage}
\hfill
\begin{minipage}[ht]{0.45\textwidth}
\begin{center}
\includegraphics[width=\textwidth,clip]{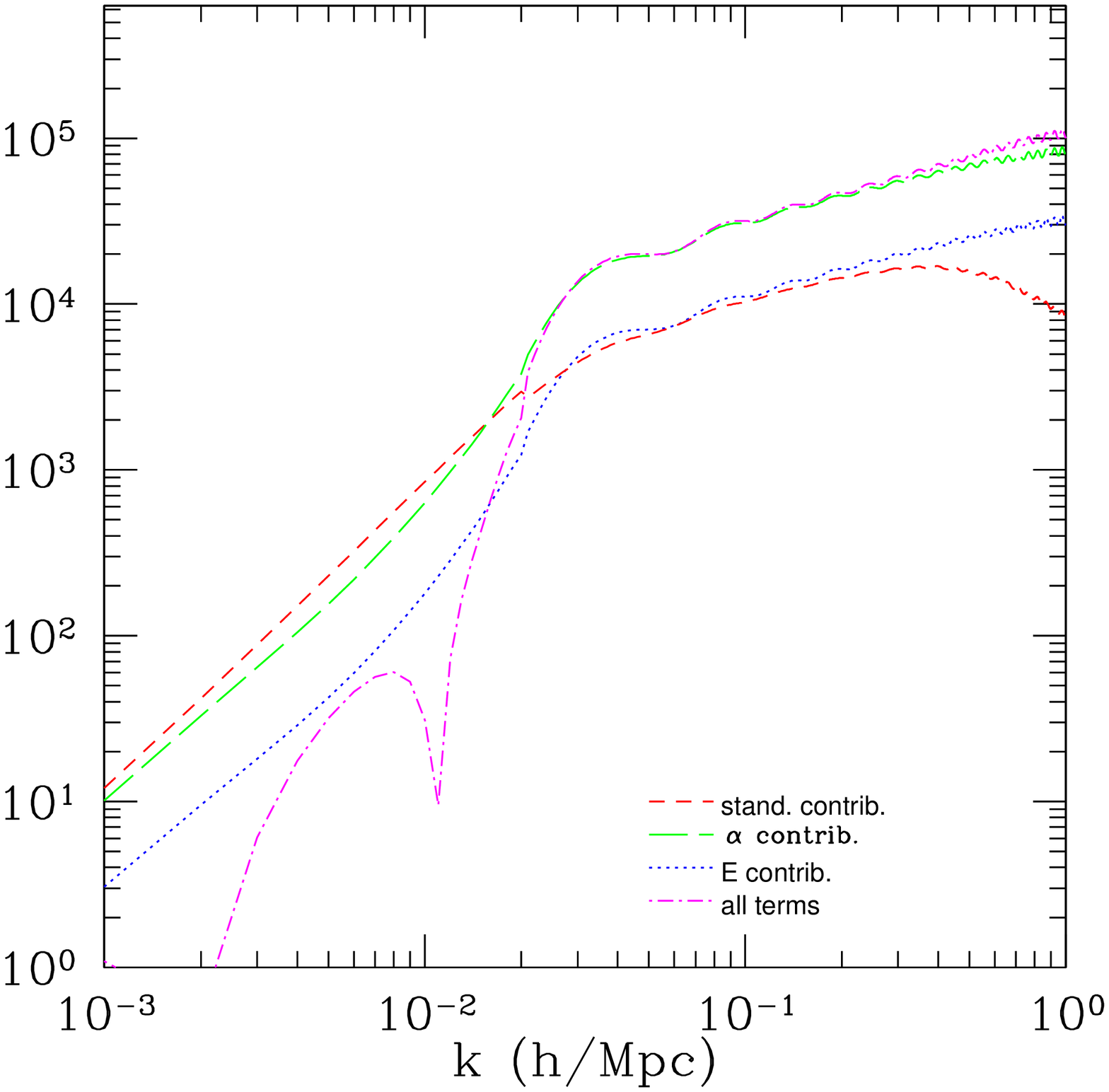}
\end{center}
\end{minipage}
\caption{\label{fig:phibaryons} {\it Left Panel.} 
$|k^2 \Phi|$ in the TeVeS baryon model (see text). After integrating 
numerically the TeVeS perturbation equations we plot the output for 
$|k^2 \Phi|$ and the different contributions to $|k^2 \Phi|$ from 
the other perturbation variables, according to 
the Poisson equation. In the plot, ``standard contrib.'' is the standard $\delta$ 
term on the R.H.S. of the Poisson equation, ``$\alpha$ contrib.''
represents
the $\left(\adotova + \phibardot \right) 
 \left(1 - e^{4 \phibar} \right)\alpha $ term and ``E contrib.'' is the
${K_B E/2}$ term (all these terms have been multiplied by $k^2$). 
``All terms'' is the sum of all the R.H.S. terms. On the x-axis $k$
is in units of h$/$Mpc but plotted lines correspond to units in
which $H_0 = 1$. Same for all the following plots.
{\it Right panel.} Same curves, but in the TeVeS neutrino model. 
}
\end{figure}

To get an equation for the other scalar potential, we use
the relation \cite{Skordis:2005eu}:
\beq\label{eq:PhiPsi}
\Psi = \Phi + e^{4 \phibar} \left[\dot{\zeta} + 2\left(\adotova +
  2\phibardot \right)\zeta \right] \; .
\eeq
From equation~(\ref{eq:zetadef}) we get:
\be
\dot{\zeta}  = -4 \phidot e^{-4 \phibar} \alpha + 
\left(e^{-4 \phibar} -1 \right) \alphadot  .
\ee
The evolution equation for $\alpha$ is:
\be
\alphadot = E + \Psi +\left(\phibardot -\adotova \right)\alpha  ,
\ee
hence:
\be
\dot{\zeta} = -4\dot{\phibar} \ex \alpha + \left(\ex-1\right)E + 
\left(\ex-1 \right)\Psi
+\left(\ex-1 \right) \left(\phibardot - \adotova \right)\alpha .
\ee
We can now include these expressions into equation~(\ref{eq:PhiPsi})
to get, after some algebra:
\beq\label{eq:PhiPsi2}
\Psi = \ex \Phi + \left (\ex-1 \right)E +
       \left[\adotova \left(\ex - 1 \right) + \phibardot \left(\ex -5
	 \right) \right] \alpha \; .
\eeq
Using the Poisson equation~(\ref{eq:poisson}) leads to:
\beq\label{eq:poissonpsi}
\Psi = -\frac{3 \ex H_0^2\left(\Omega_b \delta_b + \Omega_\nu \delta_\nu
  \right)}{2a k^2} + \left[ \ex\left(1-\frac{K_B}{2}\right)-1\right]E
-4 \phibardot \alpha \; .
\eeq
Again the first term on the right is familiar from general relativity but the vector
field perturbations induce two new source terms.
Summing $\Phi$ and $\Psi$, we obtain:
\bea
\label{eq:poissonphipsi}
\Phi + \Psi &=& -\frac{3}{2} \frac{H_0^2}{k^2}\:\left(\Omega_b \delta_b + \Omega_\nu \delta_\nu \right) (1 + \ex) + \left [ -\frac{K_B}{2} (1 + \ex)
+ \ex - 1 \right ] E - \left [ (\adotova + \dot{\bar{\phi}})(1 - \ex) 
+ 4\dot{\bar{\phi}} \right ] \alpha \nonumber\\
&\stackrel{\bar{\phi} \ll 1}{\rightarrow}&
-3\: \frac{H_0^2}{k^2} \:\left(\Omega_b \delta_b + \Omega_\nu \delta_\nu \right) - \frac{K_B}{2}\: E + 4\:\dot{\bar{\phi}}\: \alpha
\eea
The limit $\bar{\phi} \ll 1$ is imposed by nucleosynthesis
bounds on $\Omega_{\bar{\phi}}$ \cite{Skordis:2005xk}.

Equation~(\ref{eq:poissonphipsi}) can significantly 
differ from the Poisson equation in standard cosmology (\ec{GRpoisson}) 
if the vector
field perturbations $\alpha$ and $E$ are large. Two of us~\cite{Dodelson:2006zt}
showed that $\alpha$ and
$E$ get large when $K_B \rightarrow 0$. 
Thus we expect the Newtonian potentials $\Phi$ and
$\Psi$ to deviate 
from standard Poisson expectations in this strong
TeVeS regime. We also see that, unlike the standard term,
the vector field terms in
the TeVeS Poisson equation do not decrease as ${1/k^2}$. This suggests that the corrections might
actually dominate on small scales.

\begin{figure} [t]
\begin{minipage}[ht]{0.45\textwidth}
\begin{center}
\includegraphics[width=\textwidth,clip]{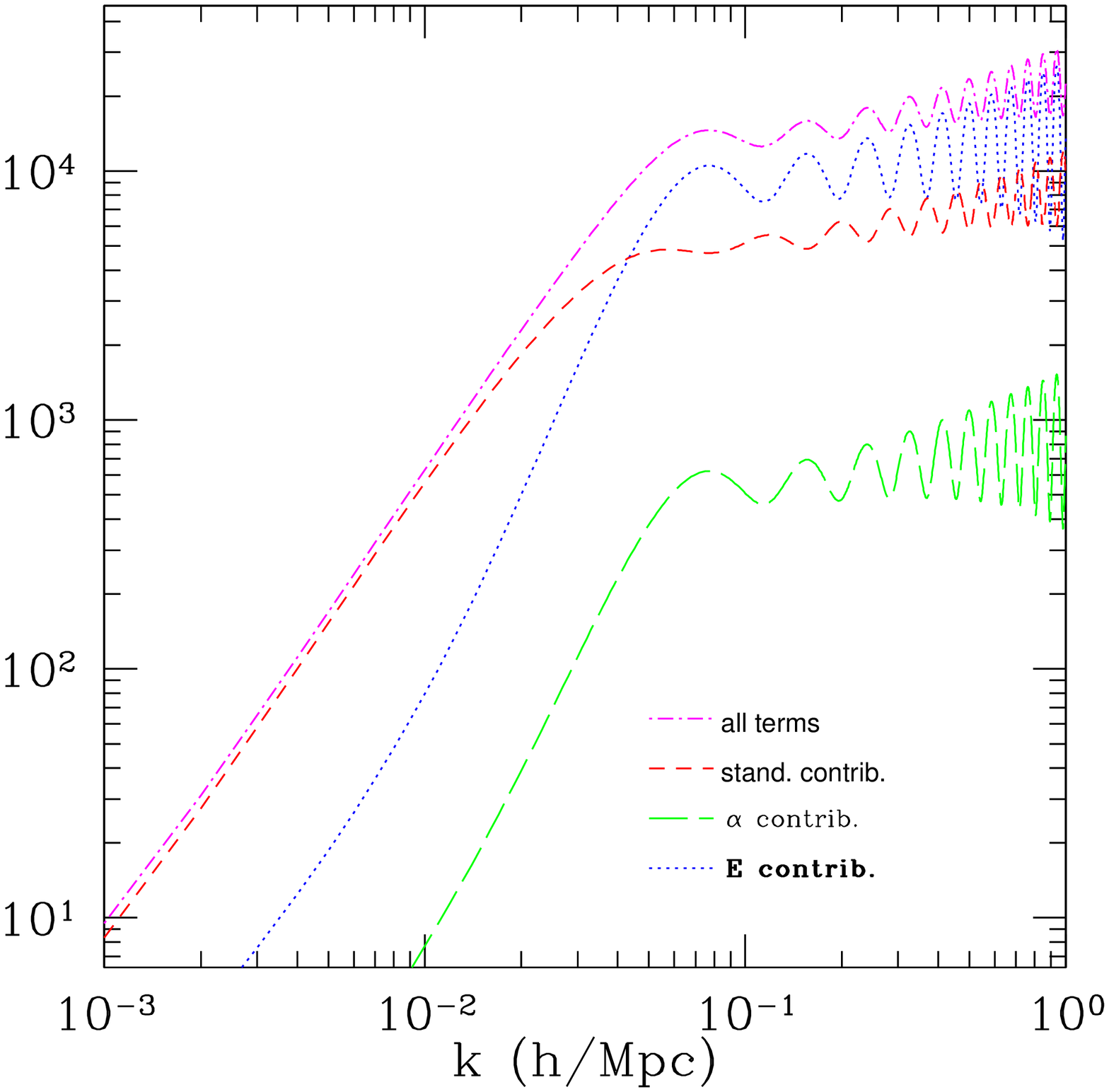}
\end{center}
\end{minipage}
\hfill
\begin{minipage}[ht]{0.45\textwidth}
\begin{center}
\includegraphics[width=\textwidth,clip]{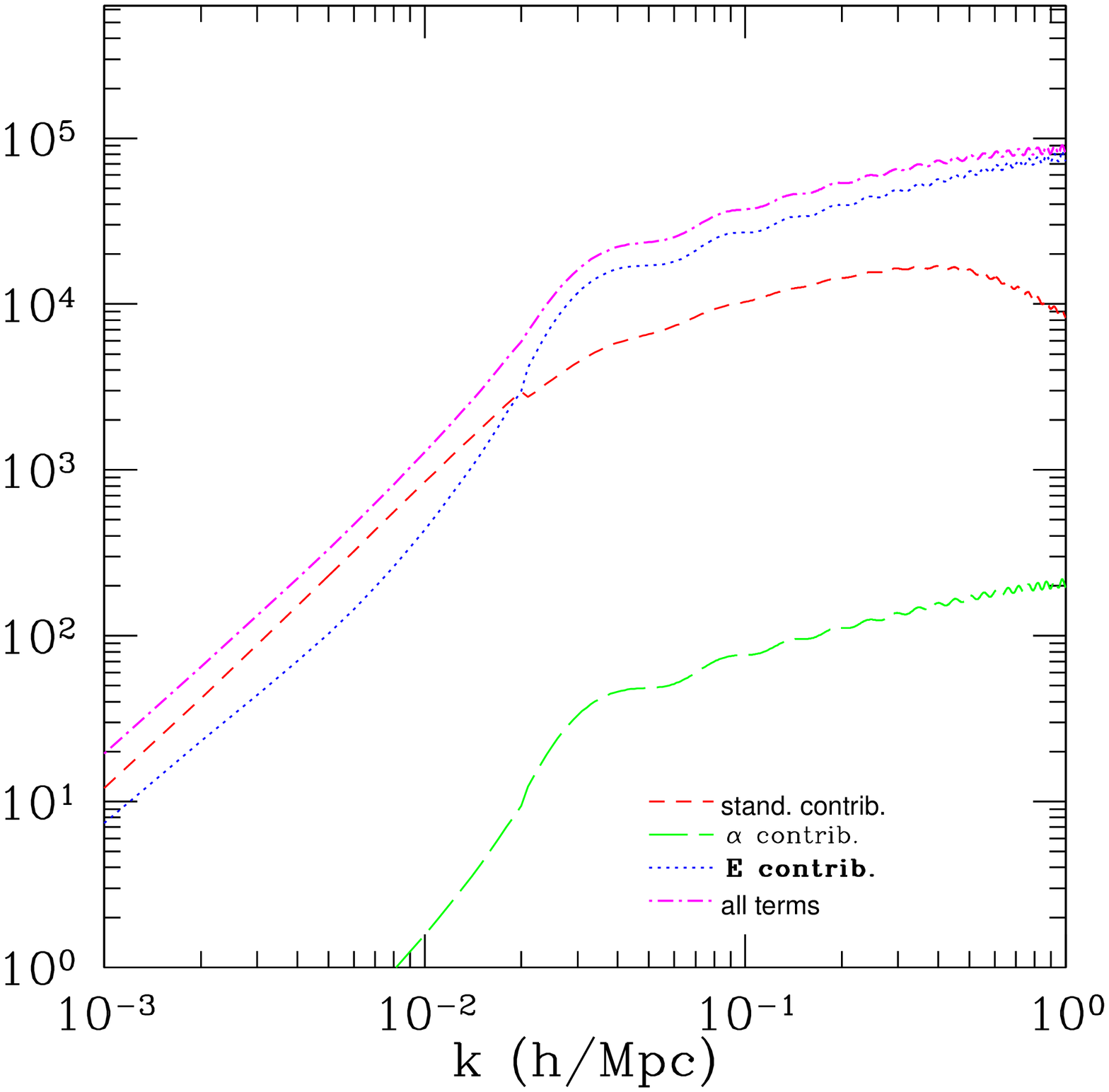}
\end{center}
\end{minipage}
\caption{\label{fig:psibaryons} {\it Left Panel.} 
$|k^2 \Psi|$ in the TeVeS baryon model. After integrating 
numerically the TeVeS perturbation equations we plot the output for 
$|k^2 \Psi|$ and the different contributions to $|k^2 \Phi|$ from 
the other perturbation variables, according to 
the equation~(\ref{eq:poissonpsi}).
{\it Right panel.} Same curves, but in the TeVeS neutrino model. 
}
\end{figure}

We solved the full set of evolution equations in two different models. The first 
(called  
``baryon model'' in the following), has only baryons with $\Omega_b =
0.3$ and $K_B = 0.08$; the second (``neutrino model'') has massive neutrinos 
with $\Omega_\nu = 0.17$, $\Omega_b = 0.05$, $K_B = 0.08$. 
The left panel of Fig.~\ref{fig:phibaryons} shows the contributions to $|k^2 \Phi|$ given by 
the different terms in equation~\ref{eq:poisson} for the baryon
model. The right panel shows the same for the neutrino
model. In both cases we can see that the standard term  
${-3 \ex H_0^2\left(\Omega_M \delta_M\right)/2ak^2}$ dominates on large
scales but it becomes basically negligible on small scales, where the 
$\alpha$ contribution is the dominant one. An analogous result is
obtained for $|k^2 \Psi|$ (Fig.~\ref{fig:psibaryons})
but this time the dominant contribution on small
scales is produced by the vector perturbation variable $E$ instead of 
$\alpha$. 

In Fig.~\ref{fig:poissonratio} 
we showed the ``Poisson ratio'' ${-k^2 (\Phi+\Psi)/(\Omega_b \delta_b +
\Omega_\nu \delta_\nu)}$. In a standard cosmological model this 
quantity is a constant but in TeVeS this is no longer 
true on small scales, where the vector field contributions in the 
Poisson equation dominate. In the baryon model the Poisson ratio shows
large oscillations on small scales. The reason for this can be
understood by looking at Fig.~\ref{fig:phibaryons} and
\ref{fig:psibaryons}. When we consider the Poisson ratio on small
scales we are basically dividing $\alpha$ and $E$ (green and blue
lines in the figures) by $\delta$ (``standard contrib.'', red line). 
We see that the
acoustic oscillations in $\alpha$ and $E$ are much more pronounced
than and out of phase with those in $\delta$. This produces the large 
oscillations that are
finally observed. Acoustic oscillations are generated by the baryon
component, which is much smaller in the neutrino model. This also 
explains why we do not see oscillations in the neutrino model Poisson
ratio. 
Another feature of the neutrino model Poisson ratio is
 a decreasing trend on small
scales (dashed-dotted blue line in Fig.~\ref{fig:poissonratio}). 
This is due to neutrinos entering the free streaming regime
and thus producing a decay in $\delta$ (red line in the right panels 
of Fig.~\ref{fig:phibaryons} and \ref{fig:psibaryons}) not matched by a
corresponding decay in $E$ and $\alpha$.

\begin{figure} [t]
\begin{minipage}[ht]{0.45\textwidth}
\begin{center}
\includegraphics[width = \textwidth]{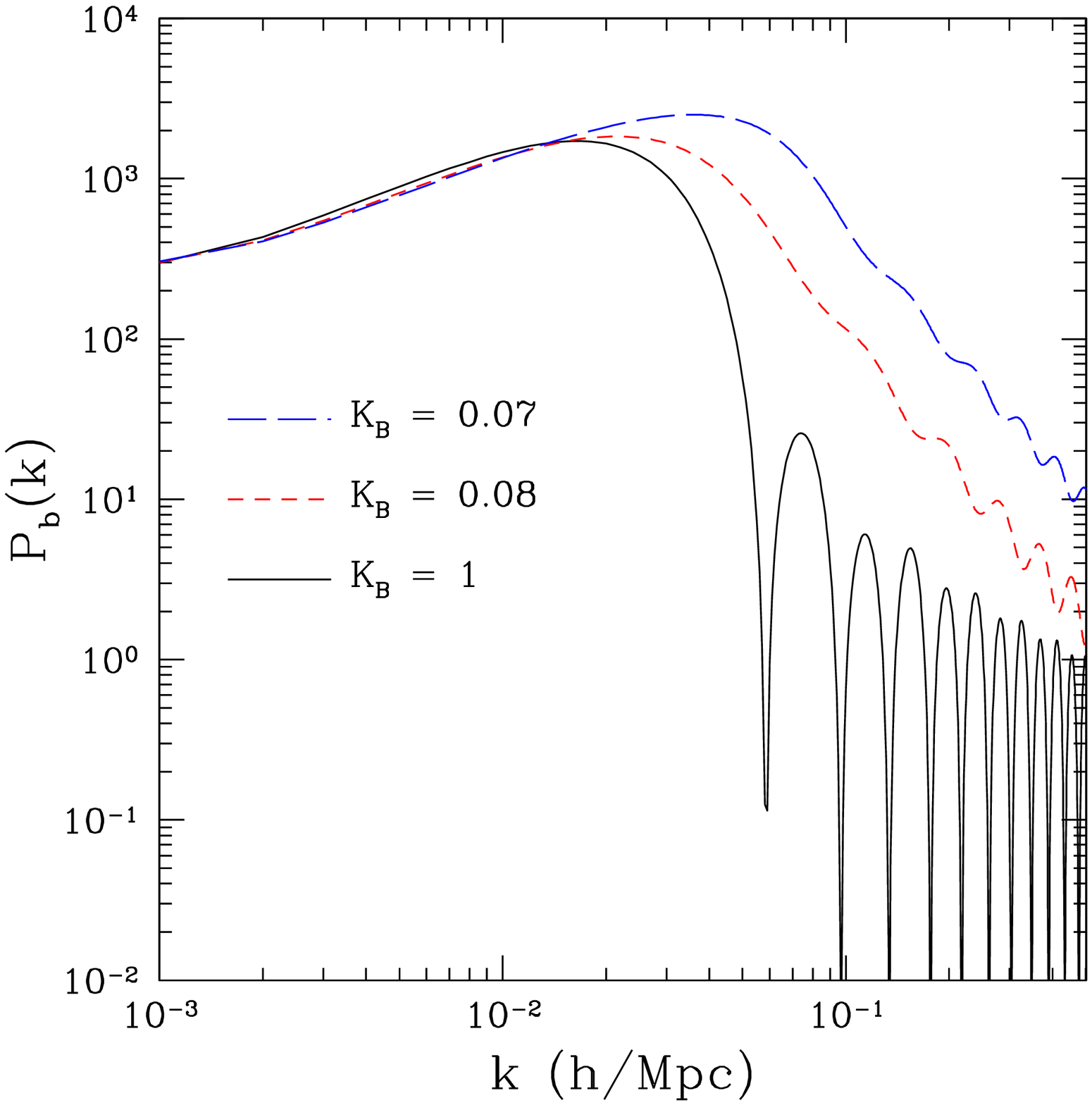}
\end{center}
\end{minipage}
\hfill
\begin{minipage}[ht]{0.45\textwidth}
\begin{center}
\includegraphics[width = \textwidth]{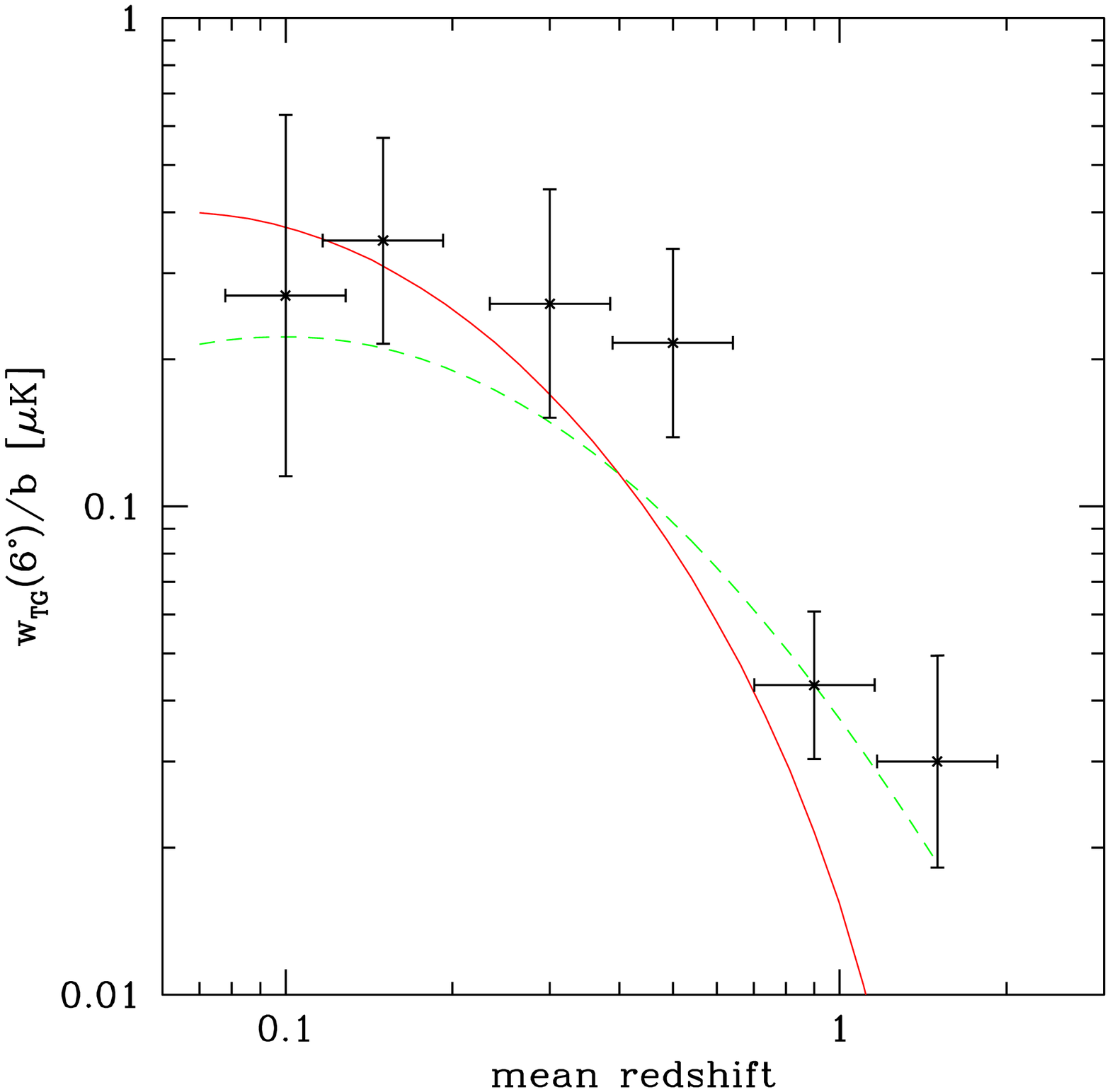}
\end{center}
\end{minipage}
\caption{{\it Left panel:} Matter power spectra for a TeVeS baryon only 
model with fixed $\Omega_b = 0.3$ and different values of $K_B$.
{\it Right panel:} Galaxy-CMB cross correlation function at $\theta=6\Deg$ 
(divided by the galaxy bias) from different surveys 
\cite{Gaztanaga,Giannantonio}.
The red (solid) curve shows the TeVeS neutrino model prediction, while
the green (dashed) curve is the $\Lambda$CDM prediction.} 
\label{fig:powerCgT}
\end{figure}   

Finally, Fig.~$\ref{fig:powerCgT}$ (left panel) shows the effect of the
vector field on the matter power spectrum. In this figure we
considered our baryon model with $\Omega_b = 0.3$ and we changed 
the value of $K_B$
while keeping the other TeVeS parameters fixed as usual. 
We know that the effect
of reducing $K_B$ is to boost the growth of vector perturbations and 
to enhance the growth factor of matter perturbations as well (see
\cite{Dodelson:2006zt}). We also see that 
reducing $K_B$ shifts the peak of the power spectrum to smaller scales.

\section{Results}\label{sec:results}

The authors of \cite{Skordis:2005xk} showed that a flat TeVeS model with
$\Omega_\nu = 0.17$, $\Omega_b = 0.05$, $\Omega_\Lambda = 0.78$
(i.e. the ``neutrino model'' we considered in the previous section)
can produce a galaxy power spectrum in reasonable agreement
with present observations. This means that a direct measurement of the power
spectrum from galaxies is not the best discriminator of the
TeVeS model and standard $\Lambda$CDM cosmology. It is then worthwhile
to ask whether there are other large scale structure
observables that can be used to distinguish between TeVeS and standard
$\Lambda$CDM, even in the case of similar predicted power spectra.

We know that in order to
match a $\Lambda$CDM power spectrum without resorting to a CDM
component, TeVeS needs a substantial contribution from the vector
field perturbations $E$ and $\alpha$. As shown in the previous
section, this leads to important modifications in the 
Poisson equation. These modifications are in turn responsible for the
change in $D_{ISW}$ with respect to standard GR illustrated 
in Fig.~\rf{iswlcdm} and can thus
potentially produce an observable effect in the galaxy-CMB cross 
correlation given by \ec{iswint} and \Ec{wgT}. Note that since the
TeVeS model contains a dark energy component, we also expect a
{\it late time} ISW effect similar to that in the standard $\Lambda$CDM 
scenario when this component starts to dominate.

For the following calculations, we use the $\Lambda$CDM linear power spectrum calculated
using the transfer function from \cite{EisensteinHu} for both TeVeS and 
$\Lambda$CDM, as it is 
accurate to the percent level, whereas the power spectrum in the TeVeS
framework is not yet known to the desired degree of precision. Throughout,
we assume the following parameters for the $\Lambda$CDM cosmology:
$h = 0.7$, $\Omega_m = 0.27$, $\Omega_{\Lambda} = 0.73$, $\Omega_b = 0.046$, 
$\sigma_8 = 0.8$, $n_s = 0.95$. 

So far, the ISW effect has been detected in the cross-correlation of the CMB
with galaxy surveys in the radio, infrared, optical, and X-ray bands
(see \cite{Gaztanaga} for a summary of recent results). These detections
span a redshift range from 0.1 to 0.9. Recently, a cross-correlation with
the SDSS quasar catalog has been performed \cite{Giannantonio} which
has a median redshift of 1.5. Fig.~\ref{fig:powerCgT} (right panel) shows the different 
observational results in terms of the angular cross-correlation function
evaluated at $\theta = 6\Deg$. Also shown are predictions from the
TeVeS neutrino model and the $\Lambda$CDM cosmology defined above. 
In order to evaluate \ec{wgT}, we assumed
a $z$-dependent galaxy selection function according to equation~(12) in
\cite{Gaztanaga}. 

Clearly, the low-$z$ (late-time) ISW effect predicted by TeVeS is quite
similar to that of $\Lambda$CDM and consistent with the current observations.
The effects of the TeVeS vector fields become noticeable only beyond $z=1$,
where there are no sufficiently strong constraints yet.

\begin{table}[b]
\center
\begin{tabular}{c|c|c|c|c}
\hline
& Bin 1 & Bin 2 & Bin 3 & Bin 4\\
\hline
$\langle z \rangle$ & 0.49 & 1.93 & 2.74 & 3.54\\
$b$ & 1.08 & 2.02 & 2.90 & 3.89\\
$n_g$ & 34.6 & 10.1 & 3.89 & 1.68\\
\hline
\end{tabular}
\caption{Parameters of the four redshift bins for the assumed LSST-like survey
\cite{LoVerde:2006cj}. $\langle z \rangle$
is the average redshift of galaxies in the bin, $b$ is the mean galaxy bias,
and $n_g$ is the observed number of galaxies per arcmin$^2$ in each 
redshift bin.}
\label{tab:param}
\end{table}

In order to determine whether future surveys will be able to 
unambiguously distinguish
the TeVeS scenario from $\Lambda$CDM, we adopt a fiducial survey 
(``Sample 1'' in Ref.~\cite{LoVerde:2006cj}) with parameters
expected from the future Large Synoptic Survey Telescope (LSST, \cite{LSST}): 
we assume
a galaxy sample with a limiting magnitude of 27 in the I band and a sky coverage of $f_s = 0.5$.
Dividing the survey into several redshift bins, the observed number
of galaxies per arcmin$^2$ as a function of redshift for each bin, $W(z)$, 
is then given by the observed $z$-dependent luminosity function of galaxies 
\cite{Gabash2004,Gabash2006} convolved
with a smoothed top hat redshift window (Eq.~(16) in \cite{LoVerde:2006cj}).
This roughly takes into account the scatter expected in the photometric
redshifts.
The mean redshifts of galaxies in the four bins considered range from 0.49 
up to 3.54. 
In order to convert from galaxy overdensity to matter 
overdensity, we have to assume a mean galaxy bias in each redshift bin.
This was calculated in \cite{LoVerde:2006cj} using the halo occupation 
distribution and the halo
mass function from simulations \cite{Kravtsov,ShethTormen}. 
In actual surveys, the bias is determined from the data itself.
The characteristics of the redshift bins are summarized in 
Tab.~\ref{tab:param}. With these parameters, we can evaluate \ec{iswint}.
Note that we neglect the magnification bias \cite{LoVerde:2006cj} as well
as redshift errors here. These effects are not expected to affect our main 
conclusions.

Fig.~\rf{CgT-MOND-LCDM} (left panel) shows the cross-power spectra for TeVeS (thick lines)
and $\Lambda$CDM (thin lines) for the different redshift
bins. As expected, the TeVeS cross-power is similar to $\Lambda$CDM in the
lowest redshift bin and goes negative at high $z$ (the ``baryon model''
considered above shows the same qualitative behavior). To estimate the
detectability of this effect, we bin the cross-power spectra in $l$,
taking the errors on a given multipole moment as:
\be
(\Delta C_{gT})^2(l) = \frac{1}{ f_s(2l+1) } \left(C_{gT}^2(l) + C_{TT}(l)  (C_{gg}(l) + \frac{1}{n_g}
 ) \right),
 \ee
where $C_{gT}$ is the galaxy-temperature cross-correlation; $C_{TT}$ and $C_{gg}$ 
are the CMB and galaxy auto-power spectra respectively; and $n_g$ is the 
number density of galaxies per sr (so that $4\pi f_s n_g$ is the total 
number of galaxies in the corresponding redshift bin of the survey; see
Tab.~\ref{tab:param}).
Throughout we set $f_s=0.5$, and errors on different $C_{gT}(l)$ are assumed
to be uncorrelated.
Fig.~\rf{CgT-MOND-LCDM} (left panel) then shows the difference between
the adopted $\Lambda$CDM model and TeVeS divided by the expected error in 
each angular bin. 
The TeVeS predictions are clearly distinguishable from 
the $\Lambda$CDM model, reaching significances of over 10$\sigma$ in the 
high redshift bins. In addition, the negative sign of the correlation
at high $z$ is an unambiguous signature of this modified gravity theory.

\begin{figure} [t]
\begin{minipage}[ht]{0.45\textwidth}
\begin{center}
\includegraphics[width=\textwidth,clip]{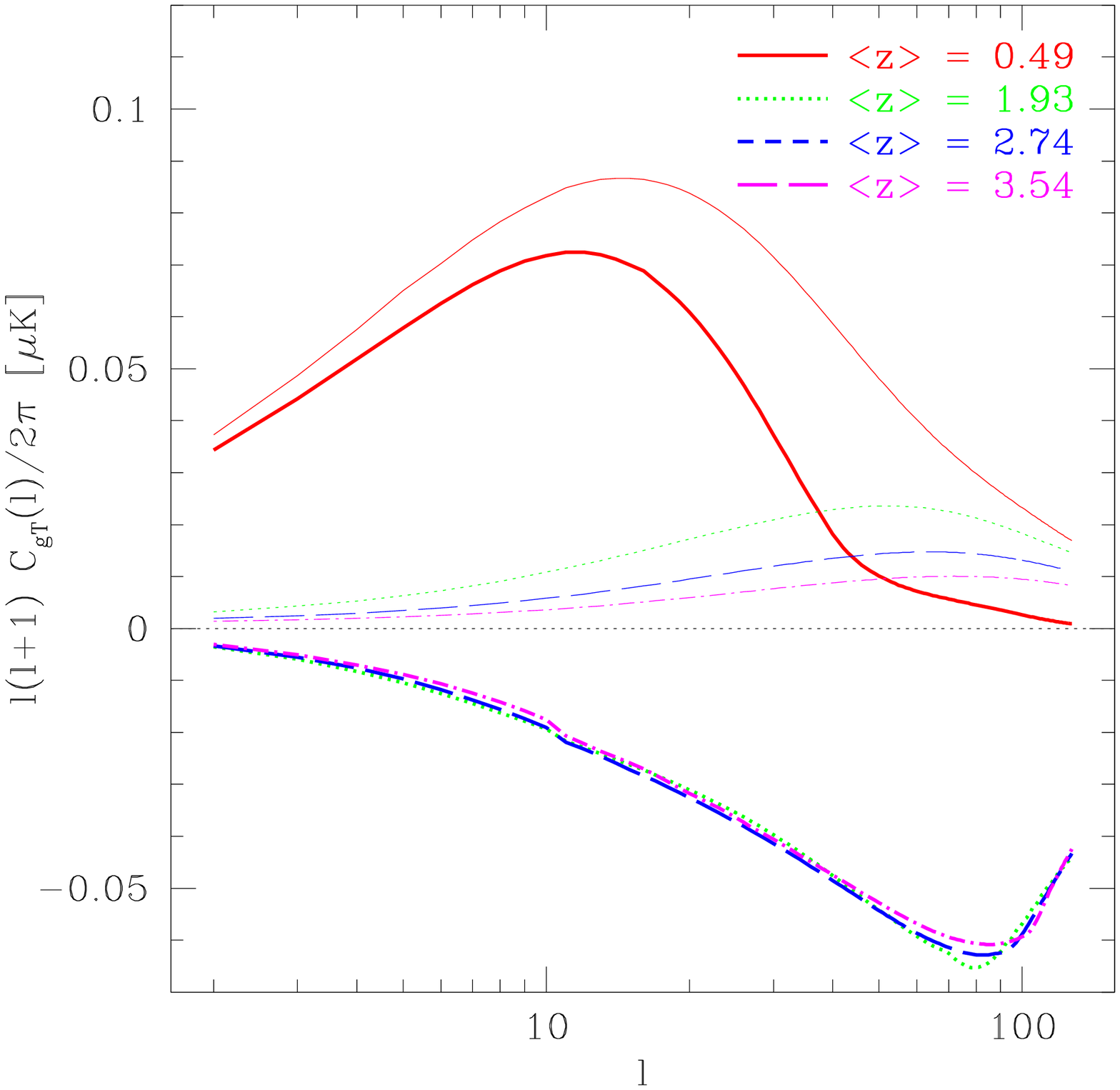}
\end{center}
\end{minipage}
\hfill
\begin{minipage}[ht]{0.45\textwidth}
\begin{center}
\includegraphics[width=\textwidth,clip]{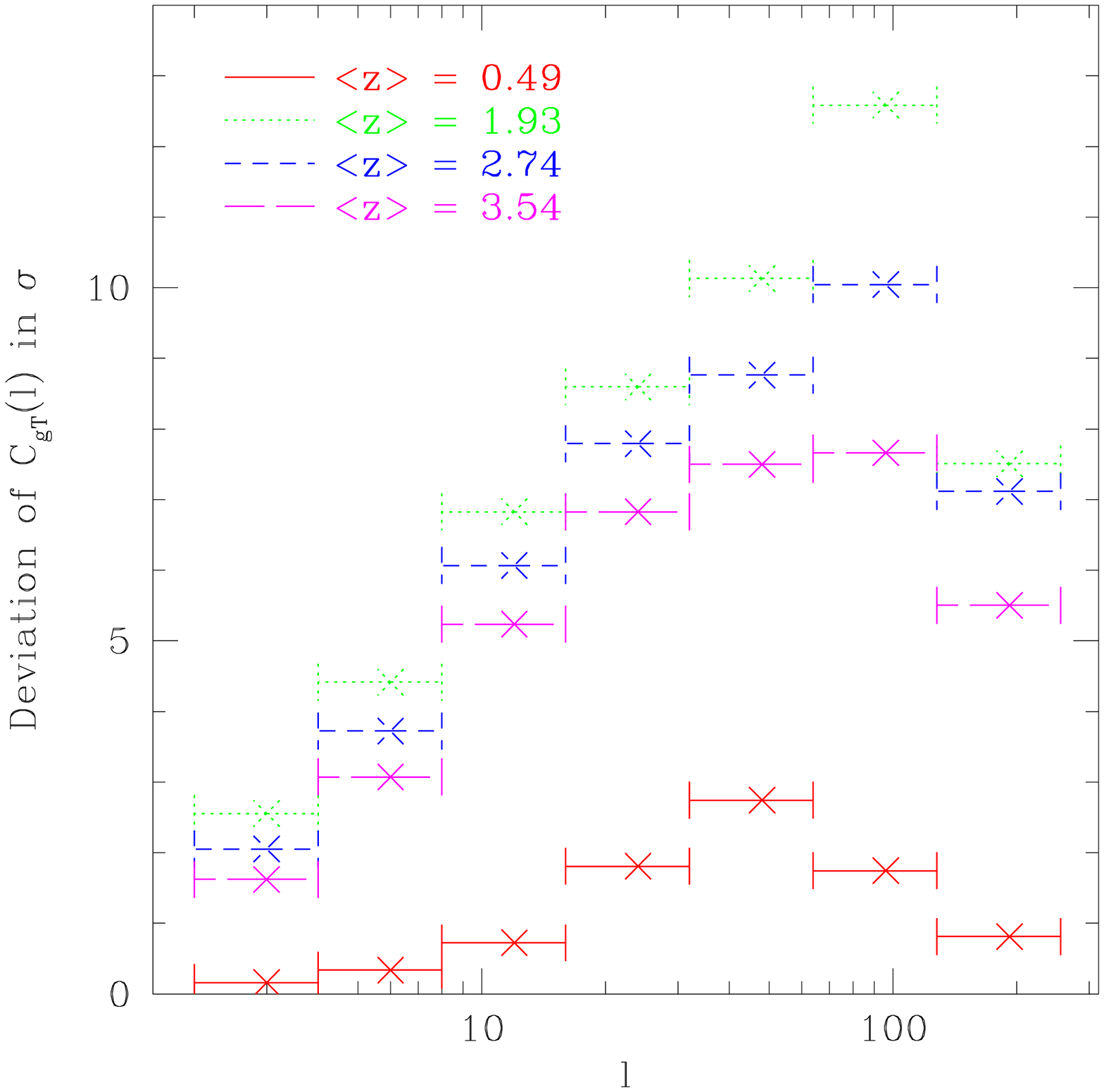}
\end{center}
\end{minipage}
\caption{{\it Left panel:} The cross-correlation of galaxies and the CMB for 
a LSST-like galaxy survey and four redshift bins, in TeVeS (thick lines) and
GR ($\Lambda$CDM, thin lines). {\it Right panel:} The significance of the 
TeVeS$-$GR deviation in the 
galaxy-CMB cross power for the LSST-like survey; i.e. difference between 
TeVeS and GR divided by the expected error in each angular bin.} 
\label{fig:CgT-MOND-LCDM}
\end{figure}   

\section{Summary and Conclusions}

TeVeS is a modified gravity theory originally introduced by 
Bekenstein~\cite{Bekenstein}
with the purpose of providing a covariant relativistic framework for
Milgrom's paradigm of MOdified Newtonian Dynamics (MOND). 
Further investigations \cite{Skordis:2005xk} showed
that a TeVeS model with massive neutrinos and no cold dark matter component can
pass some fundamental cosmological tests. In particular, it was shown
that such a model can reproduce observations of the CMB and galaxy power
spectra. 

In light of these results, in this paper we raised the following question: are there 
cosmological observables that are able to clearly distinguish between
TeVeS and standard $\Lambda$CDM ? 

In order to reproduce a $\Lambda$CDM power spectrum without
using cold dark matter, TeVeS has to incorporate a mechanism to
enhance the standard growth rate of perturbations \cite{Dodelson:2006zt}. 
This in turn produces a characteristic ISW signature
detectable in principle by looking at the CMB-galaxy cross-correlation.
In section \ref{sec:ISW} we computed the ISW effect in TeVeS,
showing that the characteristic quantity $D_{ISW}(k,z)$ presents clear
differences between TeVeS and $\Lambda$CDM at intermediate redshifts 
$z \gtrsim 1$. 
By deriving the generalized Poisson equation in the TeVeS framework
(Section \ref{sec:poisson}) we showed that the same vector field perturbations 
that are responsible for the growth of large scale structure in TeVeS are also the cause 
for the different predictions for the ISW effect.

The TeVeS model studied here is consistent with current observations of
the ISW effect (Fig.~\ref{fig:powerCgT}, right panel) which reach up to $z \approx 1.5$.
In order to determine whether the TeVeS predictions are constrainable
in future surveys, we computed
the expected CMB-galaxy cross-correlation for a fiducial survey with
parameters expected from the future LSST (Section~\ref{sec:results}). 
Our results show that at
high redshifts ($z \gtrsim 2$), TeVeS and $\Lambda$CDM will be clearly 
distinguishable, with significances of over $10 \sigma$.
Moreover we found that the sign of the CMB-galaxy cross-correlation at 
high redshifts differs between TeVeS and $\Lambda$CDM, thus providing a 
smoking-gun 
signature for this kind of modified gravity. We emphasize that this
CMB-galaxy {\it anti-correlation} at high redshifts is a robust prediction, 
since it is due to the additional degrees of freedom that are necessary
for large scale structure to form in a TeVeS cosmology.

We would finally like to stress that our work was conducted in the framework
of TeVeS because this provides a well-defined test model for which the
evolution of linear perturbations has already been studied. However,
as pointed out by Bertschinger~\cite{Bertschinger:2006aw}, the
characteristic TeVeS features that give rise to the studied effect are
expected to be common to a large class of modified gravity
models. Thus we expect the CMB-galaxy cross-correlation to be a
general and powerful tool to discriminate among many alternative
modified gravity theories.

\acknowledgments{} We would like to thank Pengjie Zhang for 
useful discussions. This work was supported by the DOE at Fermilab and 
by the Kavli Institute for Cosmological Physics at the University of Chicago 
through grants NSF PHY-0114422 and NSF PHY-0551142 and an endowment from 
the Kavli Foundation and its founder Fred Kavli. ML was supported by PPARC. 

\bibliography{submit}

\end{document}